\documentclass[aps, twocolumn, superscriptaddress, preprintnumbers]{revtex4}
\usepackage[utf8]{inputenc}
\usepackage[colorlinks, linkcolor=blue, urlcolor=blue, citecolor=blue, bookmarks, bookmarksnumbered]{hyperref}
\usepackage{bm}
\usepackage{dcolumn}
\usepackage{booktabs}
\usepackage{epstopdf}

\usepackage{xcolor}       
\usepackage{graphicx}
\usepackage{setspace}
\usepackage{subfigure}
\usepackage{multirow}
\usepackage{amssymb, amsmath}
\usepackage{siunitx} 
\usepackage{datetime} 
\usepackage{xspace}
\usepackage{hyphenat}
\usepackage{bbold}
\usepackage{afterpage}
\usepackage{multirow}
\usepackage{rotating}
\usepackage{txfonts}

\usepackage{soul}

\begin{document}

\newcommand*{\PKU}{School of Physics and State Key Laboratory of Nuclear Physics and
Technology, Peking University, Beijing 100871,
China}\affiliation{\PKU}
\newcommand*{\CHEP}{Center for High Energy Physics, Peking University, Beijing 100871, China}\affiliation{\CHEP}
\newcommand*{\CICQM}{Collaborative Innovation Center of Quantum Matter, Beijing, China}

\title{New Test of Neutrino Oscillation Coherence with Leggett-Garg Inequality\footnote{Eur.~Phys.~J.~C 82 (2022) 133\\ \url{https://doi.org/10.1140/epjc/s10052-022-10053-1}}}

\author{Xing-Zhi Wang}\affiliation{\PKU}
\author{Bo-Qiang Ma}\email{mabq@pku.edu.cn}\affiliation{\PKU}\affiliation{\CHEP}\affiliation{\CICQM}

\begin{abstract}

    Leggett-Garg inequality~(LGI) is a time analogue of Bell's inequality that concerns measurements performed on a system at different times. Violation to LGI indicates quantum coherence. We present a Leggett-Garg-type inequality compatible with more general neutrino oscillation frameworks, allowing the effects of decoherence to be taken into consideration. The inequality is applied to test coherence for data from Daya Bay, MINOS, and KamLAND experiments, and their results are compared to theoretical predictions to investigate decoherence. Both Daya Bay and MINOS data exhibit clear violations of over $10\sigma$, and of over 90$\%$ of theoretical predictions, while the KamLAND data exhibit violation of $1.9\sigma$, being of 58$\%$ of the theoretical prediction. The present work is the first to have considered the energy uncertainties in neutrino coherence tests.

\end{abstract}


\maketitle

\section{Introduction}

The idea of neutrino oscillation was proposed half a century ago~\cite{oscillation}, and has been confirmed by various sources since then. A neutrino created with a specific flavour state~$\{|\nu_{e}\rangle, |\nu_{\mu}\rangle, |\nu_{\tau}\rangle\}$ can later be found at other flavour states, and the corresponding transition and survival probabilities vary as the neutrino propagates along the space. The standard scheme of the oscillation involves 3 flavour states~$\{|\nu_{e}\rangle, |\nu_{\mu}\rangle, |\nu_{\tau}\rangle\}$ that are superpositions of mass eigenstates~$\{|m_1\rangle, |m_2\rangle, |m_3\rangle\}$~\cite{PMNS, PMNS1}.

Although most lab-generated neutrinos~(reactor neutrinos and accelerator neutrinos) exhibit oscillation behaviors that can be fitted into the standard scheme, neutrinos produced by more distant sources, such as solar activities and supernova explosions, behave rather differently. This diversity can be addressed to effects such as quantum decoherence, {\it i.e.} loss of the quantum mechanical feature named quantum coherence. Under decoherence, the flavour transitions of neutrino oscillation are underdamped and can ultimately disappear~\cite{NDP1, NDP2}. Environmental perturbation and wave-packet delocalization are generally viewed as main sources of neutrino decoherence~\cite{NDEC1, NDEC2, NDEC3, NDEC4}. While signs of wave-packet delocalization were hardly found in lab-generated neutrinos, it is suggested that environment-induced decoherence can be found in lab-generated neutrinos, especially in long-baseline experiments, and can help to explain certain results.~\cite{KD1, NDEC5, NDEC6, NDEC7}. There are also other effects that can alter the behavior of neutrinos, such as non-standard interactions~(NSI)~\cite{NSI1, NSI2, NSI3} or possible existence of sterile neutrinos~\cite{Sterile1,Sterile2,Sterile3,Sterile4}.

The idea of experimentally testing quantum mechanical features originates from Bell~\cite{Bell}. It has been shown that quantum entanglement can be experimentally identified through violation of the famous Bell's inequality~\cite{Bell, Kochen}. While Bell's inequality concerns correlation among measurements performed on spatially separated subsystems, an analogous inequality developed by Leggett and Garg, the Leggett-Garg inequality~(LGI), concerns that on one system at different times~\cite{LGI1, LGI2}. The LGI detects violation of ``macrorealism''~(MR)~\cite{LGI2}, or exhibition of coherence in the sense of quantum mechanics. The original formulation of LGI requires successive, noninvasive measurements~(NIM)~\cite{LGI2}, wherein difficulties may arise, for in quantum mechanics a measure inevitably collapses the system. Efforts on circumventing this problem are usually characterized by employing ``weak'' measurements~\cite{Semiweak1, Semiweak2, Semiweak3, Semiweak4} or constructing alternative ``testable'' inequalities~\cite{Huelga1, Huelga2, Huelga3, Lambert}. Experimental violations of LGI have been observed in various systems~\cite{Semiweak4, Vio1, Huelga2, Lambert}.

Several approaches have been employed to study experimental neutrino oscillation decoherence from different aspects, and it is proposed that LGI can be applied to test coherence in neutrino oscillation~\cite{Propose1, Propose2}.  Violations to LGI in different neutrino sources have been confirmed under a 2-state approximation~\cite{NLGI1, NLGI2}, and studies on this topic also suggested that violation can be found in 3-state neutrino oscillations in standard scheme~\cite{NLGI3}. Resent studies on neutrino oscillation coherence~\cite{NDEC1, NDEC2, NDEC3, NDEC4, KD1, KD4, NDEC5, NDEC6, NDEC7} embrace the more general Gorini–Kossakowski–Sudarshan–Lindblad~(GKSL) framework~\cite{GKSL1, GKSL2} rather that the standard 3-state unitary evolution scheme, and test compatible with this framework is therefore worth to be investigated. Also, it is possible to improve existing works by including the effect of energy uncertainties, an effect that plays an important role in experimental observations of neutrino oscillation. Methods such as analyzing the data with parameterized decoherence models~\cite{KD1, KD2, KD3,KD4} or employing quantum resource theory tools other than Bell-like inequalities~\cite{Huang, XKS}, are applied to study the same topic as well.

The present work offers a LGI for general quantum dynamical semigroup, making the test valid for more general neutrino oscillation models~(these models cover the cases of non-standard intereaction, possible sterile neutrinos, and environmental perturbation). The inequality is applied to test the neutrino oscillation data from Daya Bay, MINOS, and KamLAND experiments. Violations are observed in all three experiments. The corresponding confidence levels are estimated quantitatively, and are compared to those of theoretical predictions. It is also notable that the analysis in this work have considered the energy uncertainties of neutrino experiments, for the first time in neutrino coherence tests.

\section{Formalism}

Consider a dichotomic projection-valued measure~(PVM, quantum measure characterized by an orthocomplete set of possible outcomes) $\hat{Q}$ with realization $\pm 1$. In the language of quantum mechanics, a dichotomic PVM is characterized by two orthocomplete Hermitian projectors $\{\hat{\Pi}^{(+)}, \hat{\Pi}^{(-)}\}$ and corresponding eigenvalues $\{\lambda^{(+)}, \lambda^{(-)}\}$~(being $\pm1$ herein). The projectors correspond to seperate subspaces $\{\mathrm{H}^{(+)}, \mathrm{H}^{(-)}\}$ that form the total Hilbert space via direct sum $\mathrm{H} = \mathrm{H}^{(+)} \oplus \mathrm{H}^{(-)} $. Under eigenbasis, the projectors have the following block-diagonal form :

\begin{equation}
\begin{split}
\hat{\Pi}^{(+)} =
\left[
\begin{array}{cc}
\hat{E}_{m\times m}& \hat{O}_{m \times n}\\
 \hat{O}_{n \times m} & \hat{O}_{n\times n}\\
\end{array}
\right ];\\
\hat{\Pi}^{(-)} =
\left[
\begin{array}{cc}
\hat{O}_{m\times m}& \hat{O}_{m \times n}\\
 \hat{O}_{n \times m} & \hat{E}_{n\times n}\\
\end{array}
\right ],
\end{split}
\end{equation}
wherein $\hat{E}$ stands for identity matrix. The measure, being $\hat{Q} = \hat{\Pi}^{(+)}-\hat{\Pi}^{(-)}$, determines which subspace the system lies in, with the corresponding probabilities $P^{(+)} = \mathrm{Tr}[\hat{\Pi}^{(+)}\hat{\rho}]$ and $P^{(-)} = \mathrm{Tr}[\hat{\Pi}^{(-)}\hat{\rho}]$ and outcome $\langle Q \rangle = P^{(+)}-P^{(-)} = 2P^{(+)}-1$. The bracket $\langle\ldots\rangle$ indicates expectation, or average over many trials.

Quantum coherence, being a measure-based property characterized by nonzero off-diagonal blocks in density matrix under eigenbasis of the measure, implies that the system cannot be seen as ``in either $\mathrm{H}^{(+)}$ or $\mathrm{H}^{(-)}$~(macrorealism)''. Completely incoherent density matrices take the blockwise-diagonal form:

\begin{equation}
\hat{\rho} =
\left[
\begin{array}{cc}
 \hat{\rho}^{(+)}_{m\times m}& \hat{O}_{m \times n}\\
 \hat{O}_{n \times m} & \hat{\rho}^{(-)}_{n\times n}\\
\end{array}
\right ].
\end{equation}

Let the measure $\hat{Q}$ be performed at the system successively and noninvasively~(NIM) at different times. Defining the correlation $C(t_i, t_j)$ between time $t_i$ and $t_j$ as:

\begin{equation}
C(t_i, t_j) = \langle Q(t_i) Q(t_j) \rangle,
\end{equation}
and macrorealism predicts:

\begin{equation}
\sum_{i=1}^{N-1} C(t_i, t_i+1) - C(t_1, t_N) \leq N-2,
\end{equation}
which is the Wigner-type LGI~\cite{Wigner}. This can be easily recognized, as a single macrorealism system yields $Q = \pm 1$~(no bracket, for the result is definite) and thus satisfies the inequality. Averaging over an ensemble preserves the inequality.

Experimental determination of correlation involves successive, non-invasive measurements, which contradict the fundamental quantum mechanical principle of quantum collapse. This issue can be handled by deriving experimentally testable inequalities on the basis of additional assumptions. In particular, many researches on testing LGI in neutrino oscillations adopt the assumption of stationary correlation~\cite{NLGI1, NLGI2}:

\begin{equation}
C(t_i, t_j) = C(0, t_j-t_i),
\end{equation}
that is, the correlation depends only on time interval $t_j-t_i$, rather than on $t_i$ and $t_j$. Given the initial state, one single measurement at $t = t_j-t_i$ would be sufficient for acquiring the correlation $C(t_i, t_j)$. The issue of NIM is thus obviated.

Stationary correlation assumption holds generally true only for 2-state time-homogeneous Markovian evolution, while neutrino oscillation involves at least 3 states~($|\nu_{e}\rangle, |\nu_{\mu}\rangle, |\nu_{\tau}\rangle$). Previous studies~\cite{NLGI3} have suggested that additional correction term, evaluated using the neutrino Hamiltonian, can be added so that LGI test can be performed for 3-state neutrino oscillation. While successfully bypassing the limit of stationary correlation, the method has left space for improvement as well: The actual evolution of the system may deviate from the prediction using the presumed Hamiltonian, in this case the reliability of the evaluated correction term is limited. Also, by using a specific Hamiltonian, the evolution is assumed to be unitary. A unitary evolution prevents decoherence at first place, therefore raising question to the necessity of characterizing coherence.

Recent studies~\cite{NDEC1, NDEC2, NDEC3, NDEC4, KD1, KD4, NDEC5, NDEC6, NDEC7} on neutrino oscillation coherence embrace the GKSL approach~\cite{GKSL1, GKSL2}, a general framework that describes the time-homogeneous Markovian evolution of quantum density operator $\hat{\rho}$. Such evolutions form a quantum dynamical semigroup. Time-evolution in GKSL master equation consists of a Hamiltonian term and a dissipation term:

\begin{equation}
\frac{\partial\hat{\rho}}{\partial t} = -\frac{i}{\hbar}[\hat{\mathcal{H}}, \hat{\rho}] + \frac{1}{2}\sum_{k=1}^{N^2-1}\gamma_k( [\hat{V}_k,\hat{\rho}\hat{V}_k^{\dagger}]+[\hat{V}_k\hat{\rho},\hat{V}_k^{\dagger}]).
\end{equation}
Herein $\hat{V}_k$ are dissipative operators that are accountable for decoherence, and they form a complete basis of the $N\times N$ traceless operator space. The GKSL master equation preserves total probability $\mathrm{Tr}[\hat{\rho}]$ as well as the semigroup property. The semigroup property states that time-evolution mappings of the system $\phi_{(t_j)}$ satisfy:

\begin{equation}
\phi_{(t_j)}\phi_{(t_i)} =\phi_{(t_i+t_j)}.
\end{equation}
Time-evolution mappings are linear operators acting on the space of density operators. Time-evolution mappings can be viewed as integrals of the GKSL master equation:

\begin{equation}
\hat{\rho}(t_i+t_j) = \phi_{(t_j)}[\hat{\rho}(t_i)].
\end{equation}

If any incoherent density matrix remains incoherent in its subsequent evolution, then the time-evolution mappings can be decomposed into survival and transition terms:

\begin{equation}
\begin{split}
\hat{\rho}^{(+)}(t_i+t_j) = \phi^{++}_{(t_j)}[\hat{\rho}^{(+)}(t_i)]+\phi^{+-}_{(t_j)}[\hat{\rho}^{(-)}(t_i)],\\
\hat{\rho}^{(-)}(t_i+t_j) = \phi^{-+}_{(t_j)}[\hat{\rho}^{(+)}(t_i)]+\phi^{--}_{(t_j)}[\hat{\rho}^{(-)}(t_i)].
\end{split}
\label{eq:evo}
\end{equation}

Consider a system that starts as an equiprobable distribution in $\mathrm{H}^{(+)}$ at $t = 0$. The corresponding density matrix $\hat{\rho}(0)$ is indeed incoherent, explicitly:

\begin{equation}
\begin{split}
\hat{\rho}^{(+)}(0) = \frac{1}{m}\hat{E}_{m\times m},\\
\hat{\rho}^{(-)}(0) = \hat{O}_{n\times n}.
\end{split}
\end{equation}

Perform the measure at $(t, 2t, ...)$, on separate members of an ensemble characterized by the density matrix. The NIM is circulated, for successive measurements are not involved. Density matrices and corresponding probabilities can be obtained by applying Eq.~(\ref{eq:evo}). The following inequalities can be acquired by dropping the transition terms, since they give nonnegative contribution to the survival rate $P^{(+)}$:

\begin{equation}
\begin{split}
P^{(+)}(2t) \geq \mathrm{Tr}[{\phi^{++}_{(t)}}^2[\frac{1}{m}\hat{E}_{m\times m}]]\geq P^{(+)}(t)^2,\\
P^{(+)}(3t) \geq \mathrm{Tr}[{\phi^{++}_{(t)}}^3[\frac{1}{m}\hat{E}_{m\times m}]]\geq P^{(+)}(t)^3,\\
\ldots
\end{split}
\end{equation}
The second ``$\geq$" is provided by the AM-GM inequality of the eigenvalues of density matrix. For the special case of $m = 1$, the second ``$\geq$" is replaced by ``$=$", as $\hat{\rho}^{(+)}$ is now $1\times 1$ with value $P^{(+)}$. Also, the time intervals $t$ are not necessarily identical under this circumstance. Hence, the inequality becomes:

\begin{equation}
P^{(+)}(\sum_{i=1}^{N}t_i) - \prod_{i=1}^{N} P^{(+)}(t_i) \geq 0.
\label{eq:ineq}
\end{equation}
This inequality is to be applied to test neutrino coherence in this work.

In addition, if $n = m = 1$ the contribution of transition terms can be explicitly evaluated rather than simply dropped off, resulting in the equality:

\begin{equation}
(2P^{(+)}(\sum_{i=1}^{N}t_i)-1) = \prod_{i=1}^{N} (2P^{(+)}(t_i)-1),
\end{equation}
which recovers the Wigner-type LGI for stationary correlation~\cite{Wigner}:

\begin{equation}
\sum_{i=1}^{N-1} C(0, t_{i}) - C(0, \sum_{i=1}^{N-1}t_{i}) \leq N-2.
\end{equation}

\section{Neutrino Oscillation}

The standard scheme of neutrino oscillation is the 3-state neutrino model. The oscillation involves 3 flavour eigenstates $\{|\nu_{e}\rangle, |\nu_{\mu}\rangle, |\nu_{\tau}\rangle\}$, being superpositions of mass eigenstates $\{|m_1\rangle, |m_2\rangle, |m_3\rangle\}$:

\begin{equation}
|\nu_{\alpha}\rangle = \sum_{k} U^{*}_{\alpha k} |\nu_{k}\rangle,
\end{equation}
herein $\alpha=\{e, \mu, \tau\}$, and $k=\{1, 2, 3\}$. $U_{\alpha k}$ is the Pontecorvo-Maki-Nakagawa-Sakata~(PMNS) matrix~\cite{PMNS, PMNS1}, parameterized by three mixing angles and one CP-violating phase:

\begin{equation}
\left(
           \begin{array}{ccc}
             c_{12}c_{13} &  s_{12}c_{13} &  s_{13}e^{-i\delta_{CP}} \\
             -\!s_{12}c_{23}\!-\!c_{12}s_{13}s_{23}e^{i\delta_{CP}} & c_{12}c_{23}\!-\!s_{12}s_{13}s_{23}e^{i\delta_{CP}} & c_{13}s_{23}\\
             s_{12}s_{23}\!-\!c_{12}s_{13}c_{23}e^{i\delta_{CP}} & -\!c_{12}s_{23}\!-\!s_{12}s_{13}c_{23}e^{i\delta_{CP}} & c_{13}c_{23} \\
           \end{array}
         \right).
\end{equation}

Neutrino in vacuum at ultrarelativistic limit subjects to a unitary time-evolution determined by the PMNS matrix and two mass square differences $\Delta m^2_{21}, \Delta m^2_{31}$, if decoherence effects are not taken into account. The survival and transition rates under the standard scheme are given as:

\begin{equation}
\begin{split}
P_{\alpha\rightarrow\beta} = |\sum_{i} U^{*}_{\alpha i} U_{\beta i} e^{-i\frac{\Delta m^2_{i1} L}{2E}}|^2.
\end{split}
\label{eq:osc}
\end{equation}
The parameter $L/E$ herein can be viewed as an analogue of ``time". In the following section Eq.~(\ref{eq:osc}) provides theoretical prediction of survival rates. The oscillation parameters are listed in Table 1.

\begin{table}[htb]
      \caption{Neutrino oscillation parameters~(Normal Ordering) from global fit. The data are taken from NuFIT~\cite{NFIT, NFIT1}. }
      \label{tab:1}
      \begin{ruledtabular}
      \centering
      \begin{tabular}{ccc}

       parameter & best fit$\pm 1\sigma$\  & $3\sigma$ range\\
      \hline
      \noalign{\vspace{0.5ex}}
      $\Delta m^2_{21}/10^{-5}~{\rm eV^2}$ & $7.42_{-0.20}^{+0.21}$ & $6.82 \to 8.04$\\
      \noalign{\vspace{0.5ex}}
      $\Delta m^2_{31}/10^{-3}~{\rm eV^2}$ & $2.514_{-0.027}^{+0.028}$ & $2.431 \to 2.598$\\
      \noalign{\vspace{0.5ex}}
      $\theta_{12}/^\circ$ & $33.44_{-0.75}^{+0.78}$ & $31.27 \to 35.86$ \\
      \noalign{\vspace{0.5ex}}
      $\theta_{13}/^\circ$ & $8.57_{-0.13}^{+0.13}$ & $8.20 \to 8.97$ \\
      \noalign{\vspace{0.5ex}}
      $\theta_{23}/^\circ$ & $49.0_{-1.4}^{+1.1}$ & $39.6 \to 51.8$ \\
      \noalign{\vspace{0.5ex}}
      $\delta_{CP}/^\circ$ & $195_{-25}^{+51}$ & $107 \to 403$ \\

      \end{tabular}
      \end{ruledtabular}
\end{table}

The general GKSL framework allows the inclusion of effects beyond the standard scheme, such as non-standard interaction, sterile neutrinos, and environmental perturbation. These effects modify the evolution Eq.~(\ref{eq:osc}) by altering the Hamiltonian, adding new flavour eigenstates to the Hilbert space, and introducing additional non-unitary dissipation terms. The inequality Eq.~(\ref{eq:ineq}) is deduced using this framework, and is therefore applicable for coherence test even with these effects in presence.

\section{Experimental Violation of LGI}

In experiments, neutrinos with almost identical initial state~($|\nu_{e}\rangle, |\nu_{\mu}\rangle, |\nu_{\tau}\rangle$) travel across a fixed baseline $L$ before reaching the detector. Changing the energy $E$ of the neutrinos results in different survival(~or transition) probabilities $P_{\mathrm{suv}}$ and their dependance on $L/E$.

\begin{widetext}
	\begin{figure}[ht]
	\centering
	\begin{tabular}{ccc}
		\includegraphics[width=60mm]{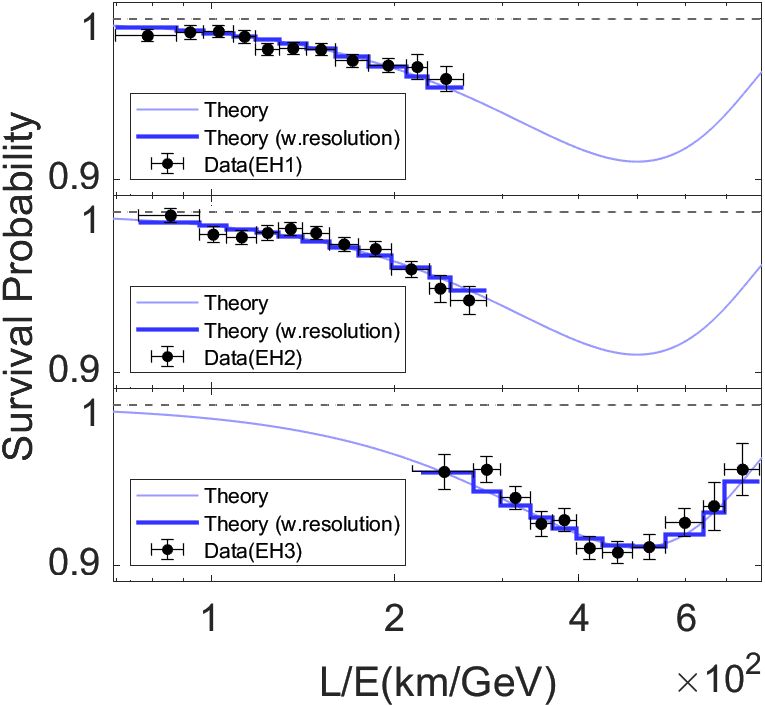}
		\includegraphics[width=60mm]{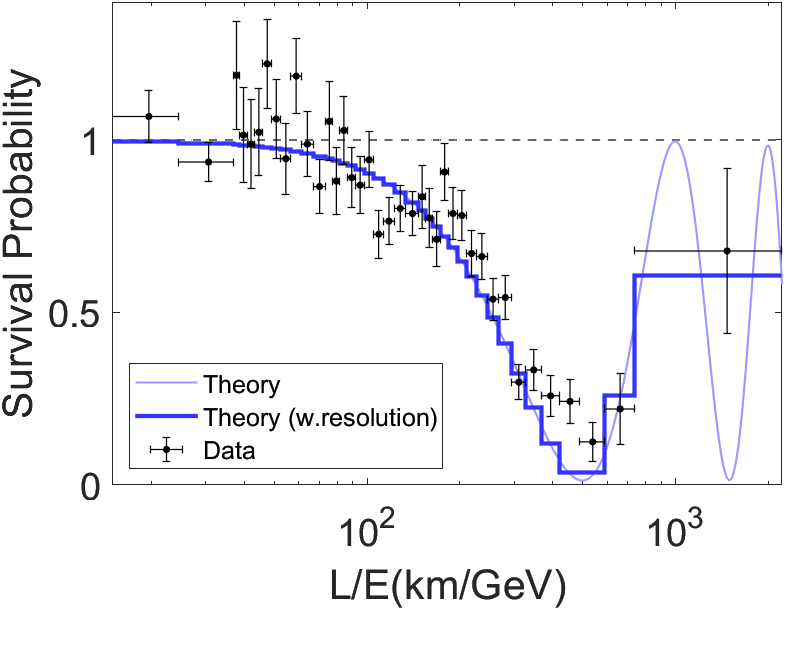}
		\includegraphics[width=60mm]{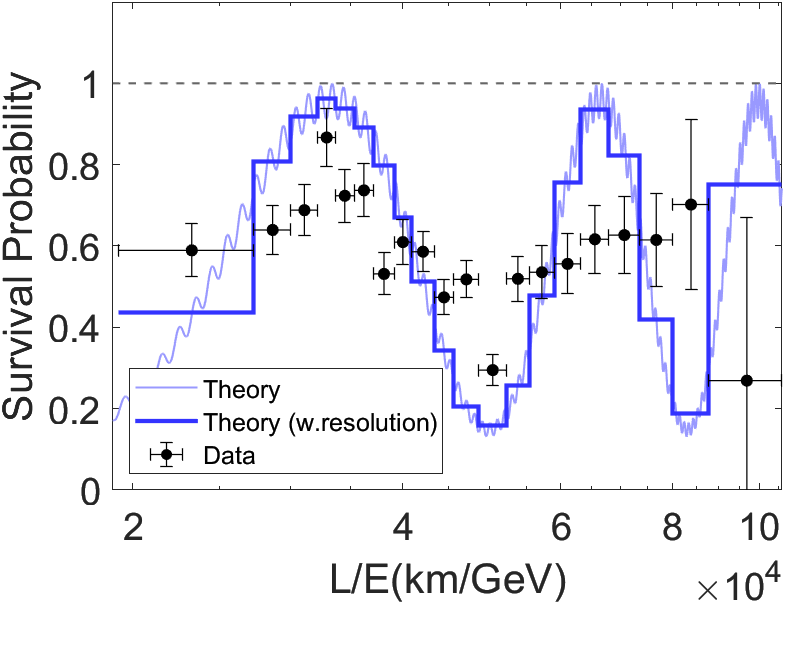}
	\end{tabular}
    \caption{ Neutrino survival probability data from Daya Bay~(left)~\cite{DayaBay2}, MINOS~(middle)~\cite{MINOS2}, and KamLAND~(right)~\cite{KamLAND2}. Blue curves indicate the theoretical prediction of standard scheme Eq.~(\ref{eq:osc}) using parameters from NuFIT~\cite{NFIT, NFIT1} global fit. The stairs represent the predicted value averaged over the uncertainty interval of corresponding data points, demonstrating the `flattening' effect that arises from the finite energy resolution of the experiments. Data from the three experiment halls~(EHs) of Daya Bay~(left) are demonstrated separately.}
	\label{Psuv}
\end{figure}
\end{widetext}

The work tests coherence for data gathered by MINOS, Daya Bay, and KamLAND with LGI. All the three experiments are disappearance experiments that measure neutrino survival rates. Both Daya Bay and KamLAND study electron-antineutrino~($|{\bar{\nu}}_{e}\rangle$), and MINOS studies muon-neutrino~($|\nu_{\mu}\rangle$). As for the baseline and energy parameters, Daya Bay has $L\in [364\mathrm{m}, 1912\mathrm{m}]$~(for there are multiple reactors and detectors located differently) and $E\in[1\mathrm{MeV}, 8\mathrm{MeV}]$~\cite{DayaBay1}, MINOS has $L = 735\mathrm{km}$ and $E \in [0.5\mathrm{GeV}, 50\mathrm{GeV}]$~\cite{MINOS1}, and KamLAND has $L = 180\mathrm{km}$ and $E\in[2\mathrm{MeV}, 10\mathrm{MeV}]$~\cite{KamLAND1}. The survival rates obtained by these experiments and theoretical predictions of 3-neutrino model Eq.~(\ref{eq:osc}) are shown in Fig.~\ref{Psuv}

\onecolumngrid

For neutrino disappearance experiments, Eq.~(\ref{eq:ineq}) is applicable, with the measure projectors being $\hat{\Pi}^{(+)} = |i\rangle \langle i |$ and $\hat{\Pi}^{(-)}=\hat{E}-|i\rangle \langle i |$, wherein $|i\rangle$ stands for the initial state. The $3$-party inequality becomes:

\begin{equation}
K_{3}(t_i, t_j, t_k) := P_{\mathrm{suv}}(t_i)P_{\mathrm{suv}}(t_j) - P_{\mathrm{suv}}(t_k)  \leq 0,
\label{eq:ineq3}
\end{equation}
with
\begin{equation}
t_i+t_j=t_k.
\label{eq:correlate}
\end{equation}

For real data, however, the `time' $t=L/E$ has systematic uncertainties~(for that the energy resolutions are finite), and can therefore never match exactly the correlation condition of Eq.~(\ref{eq:correlate}). The observed survival probabilities have uncertainties as well. To take these uncertainties into consideration as well as to evaluate qualitatively the degree of LGI violation, the method of statistical sampling of generated pseudodata can be utilized.

The procedure goes as follows: Given a set of experimental data of times and survival rates and corresponding uncertainties $\{(t_i, P_i; \Delta t_i, \Delta P_i)\}$, generate a set of pseudodata of times and survival rates $\{(\tau_i, \pi_i)\}$ using normal distributions, with the means being $(t_i, P_i)$ and the variances being $((\Delta t_i)^2, (\Delta P_i)^2)$.

A data triad $\{(\tau_i, \pi_i), (\tau_j, \pi_j), (\tau_k, \pi_k)\}$ is taken as ``correlated'' if it satisfies the correlation condition of Eq.~(\ref{eq:correlate}) within the range of $\epsilon = 5\%$:

\begin{equation}
\frac{|\tau_i + \tau_j - \tau_k|}{\tau_k} \leq \epsilon.
\label{eq:ineq3}
\end{equation}

The reason for setting $\epsilon$ to be $5\%$ is that, if the standard is too strict~(for example, $\epsilon = 1\%$), it would be virtually impossible to have any correlated triads for the time uncertainties of the data being considered herein. On the other hand, a too rough standard, such as $\epsilon = 10\%$, would hardly resemble any actual correlation. Setting $\epsilon$ to be $5\%$ strikes a balance between these two factors, and it is also a conventional value in statistics.

A correlated triad is considered to have violated LGI, if:

\begin{equation}
K_{3}(\tau_i, \tau_j, \tau_k) = \pi_i \pi_j - \pi_k  \textgreater  0.
\label{eq:ineq3}
\end{equation}

Count the number of correlated triads that exhibit violations. Repeating this procedure generates a distribution of violation counts. Confidence level of the violation can be obtained by dividing the expectation of the distribution $\mu$ by its standard deviation $\sigma$ .

With this data processing procedure, both the MINOS and the Daya Bay neutrino oscillation data yield a clear violation of over $10\sigma$, while the KamLAND data exhibit a less significant violation, of $1.9\sigma$. However, these results cannot be compared directly, as more data points and better energy resolution would result in higher confidence level even for the same evolution.

To understand the implication of these results, the same procedure is applied for data with experimental values of times, time uncertainties, and survival rate uncertainties, while the survival rates are replaced by theoretically predicted values Eq.~(\ref{eq:osc}), and the results are compared with those of experimental data. The confidence levels of violation of both Daya Bay and MINOS data are very close to those of theoretical pseudodata, of over 90$\%$, while that of KamLAND data is considerably less than the confidence level of theoretical pseudodata, of only 58$\%$. The results and the comparison are shown in Fig.~\ref{Violations}, and the generated pseudodata are shown in Fig.~\ref{K3}.

\onecolumngrid
\begin{widetext}
	\begin{figure}[ht]
	\centering
	\begin{tabular}{ccc}
		\includegraphics[width=60mm]{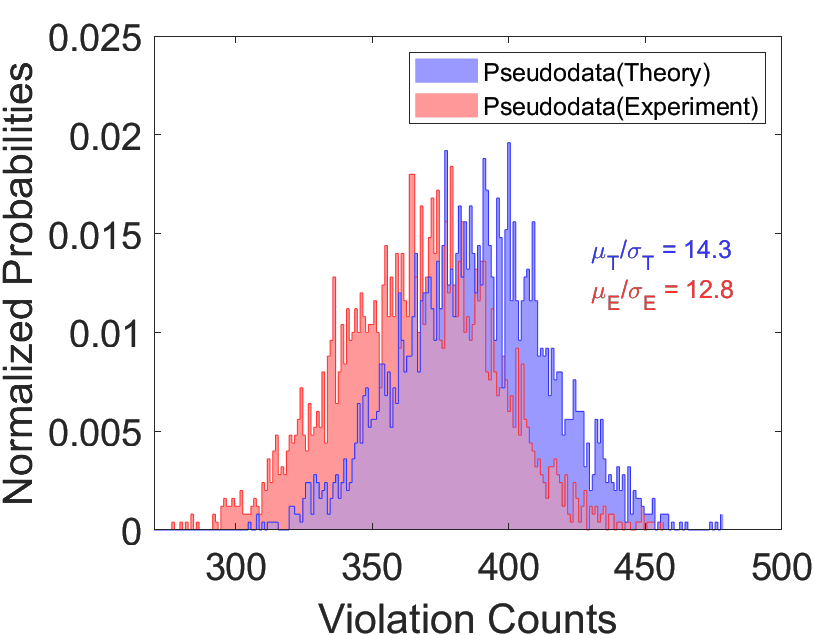}
		\includegraphics[width=60mm]{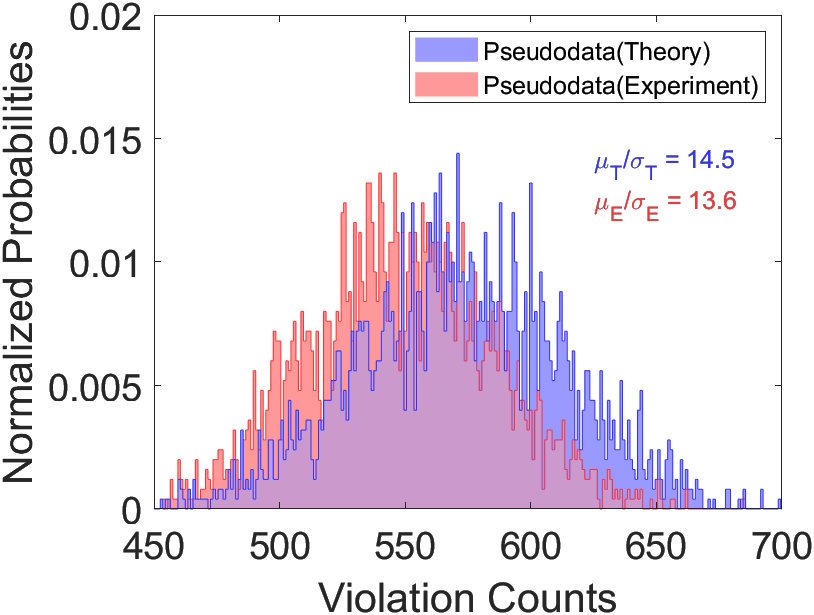}
		\includegraphics[width=60mm]{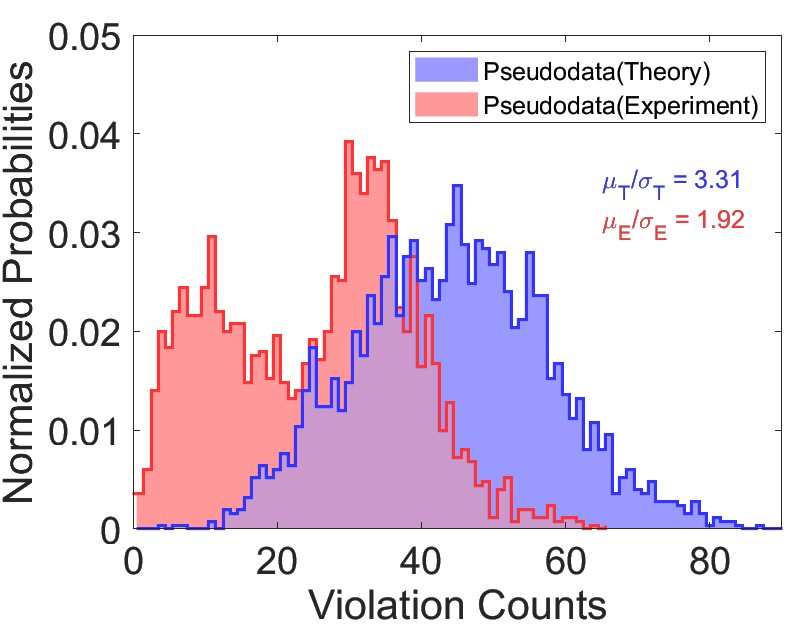}
	\end{tabular}
	\caption{ Statistics of violation counts of experimental and theoretical pseudodata for Daya Bay~(left)~\cite{DayaBay2}, MINOS~(middle)~\cite{MINOS2}, and KamLAND~(right)~\cite{KamLAND2}. The distributions and violation confidence levels of Daya Bay and MINOS experimental pseudodata are close to that of theoretical pseudodata~(90$\%$ and 94$\%$), while the distribution of KamLAND experimental data is very different from its theoretical pseudodata, and the violation confidence level is only 58$\%$ of theoretical pseudodata. } 
	\label{Violations}
\end{figure}

\onecolumngrid
\begin{widetext}
	\begin{figure}[ht]
	\centering
	\begin{tabular}{ccc}
		\includegraphics[width=60mm]{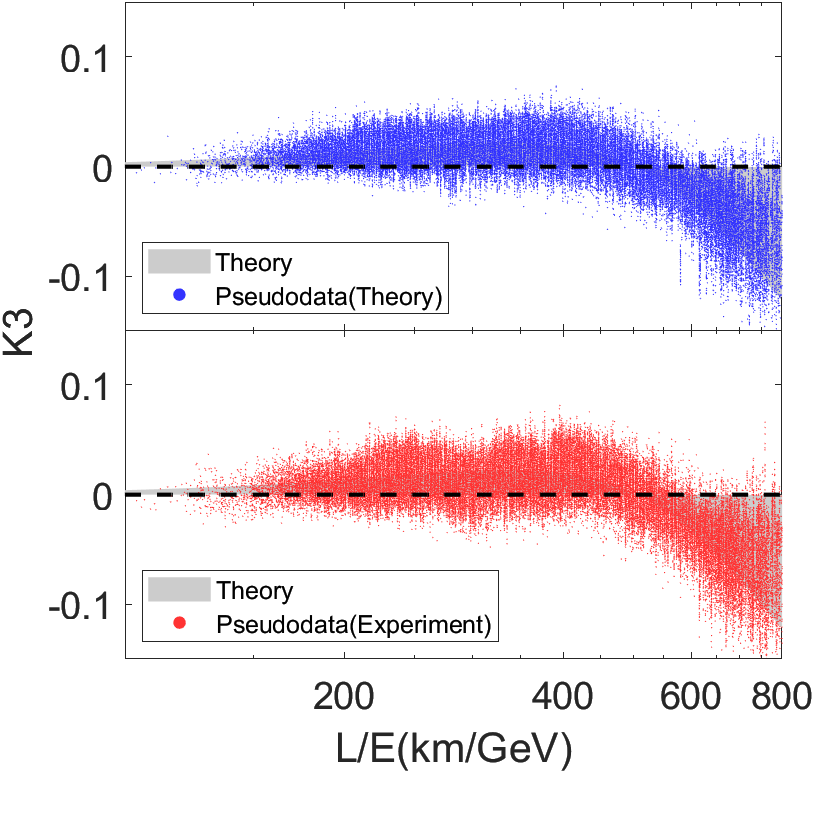}
		\includegraphics[width=60mm]{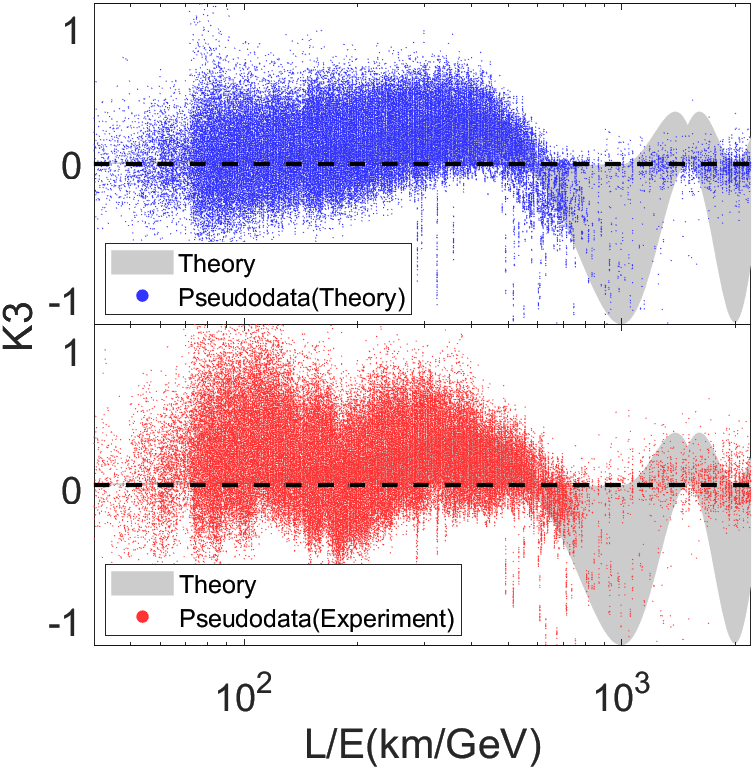}
		\includegraphics[width=60mm]{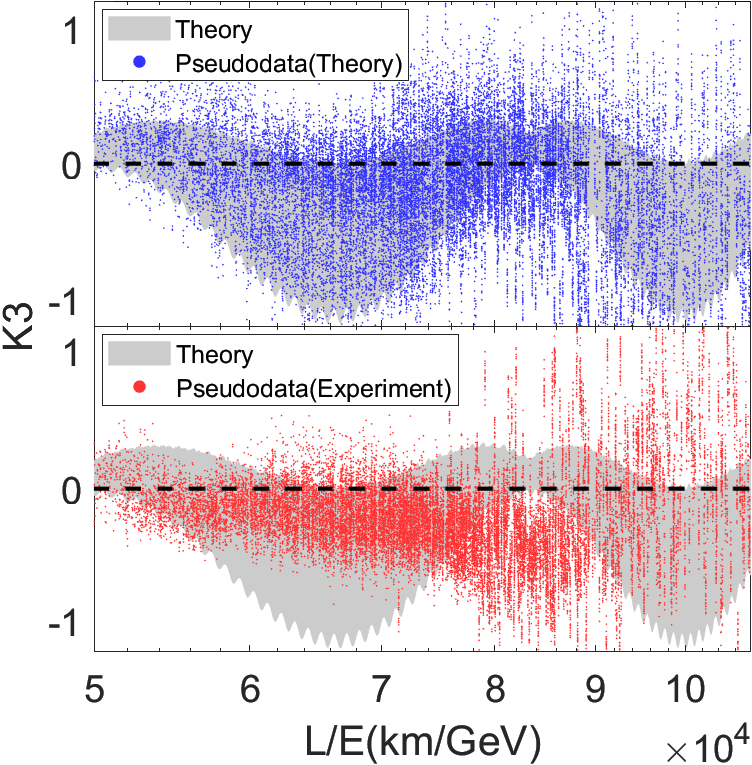}
	\end{tabular}
	\caption{ Distributions of pseudodata of $K_3$ for Daya Bay~(left)~\cite{DayaBay2}, MINOS~(middle)~\cite{MINOS2}, and KamLAND~(right)~\cite{KamLAND2}. The gray shade represent the prediction using standard scheme Eq.~(\ref{eq:osc}) with NuFIT~\cite{NFIT, NFIT1} global fit parameters, while the blue and red points represent experimental and theoretical pesudodata separately. }
	\label{K3}
\end{figure}
\end{widetext}

\end{widetext}

\section{Conclusion}

This work presents a testable Leggett-Garg-type inequality compatible with recent GKSL framework of neutrino oscillation study. The inequality allows test for a wide range of neutrino oscillation models~(such may involve environmental perturbation, non-standard interactions, and possible sterile neutrinos), and have the potential of identifying decoherence. This work advances existing works in this area by offering a general test that is free from 2-state approximation or unitary~(no decoherence) evolution assumption. The analysis in this work have also taken the experimental energy uncertainties into account, being the first time in neutrino coherence tests.

The inequality is applied to test coherence for neutrino oscillation data from MINOS, Daya Bay, and KamLAND. Both MINOS and Daya Bay data give definite results of LGI violation, while the violation in KamLAND data is not as clear. The results are compared to those of theoretical prediction. Violation confidence levels of Daya Bay and MINOS are close to those of theoretical pseudodata~(90$\%$ and 94$\%$), while the confidence level of KamLAND is only 58$\%$ of that of theoretical pseudodata. KamLAND has one of the biggest ``time" parameter ($L/E \sim 10^{5}  $km/GeV) among current neutrino oscillation experiment facilities~\cite{Rayner} and its data deviate from the prediction of the standard 3-flavour unitary evolution with global fit parameters, and KamLAND data therefore have been used as a source for investigating neutrino decoherence~\cite{KD1, KD2, KD3}. The result in this work supports the idea of seeking decoherence in KamLAND and other lab-generated neutrino experiments, for the confidence level of LGI violation in KamLAND data is considerably less than that of theoretical prediction.

The question of testing neutrino coherence with the presence of wave-packet delocalization, is still unsolved. The delocalization effect is characterized by the parameter $L^2/E^4$~\cite{NDP1, KD3}, and thus a universal ``time" parameter with respect to energy $E$ and baseline $L$ can no longer be constructed. Besides, the effective time-evolution of delocalization in flavour subspace goes beyond the GKSL scheme for it is non-Markovian. Further efforts are needed to tackle these difficulties. Also, it is still yet to de determined that what mechanisms lead to the difference between KamLAND data and that of theoretical prediction, and that how to describe the coherence loss over propagation in experiments quantitatively.

\begin{acknowledgments}
This work is supported by National Natural Science Foundation of China (Grants No.~12075003) and by the President's Undergraduate Research Fellowship (PURF) of Peking University.

\end{acknowledgments}

\end{document}